# Multiplexed vortex beam-based optical tweezers

Francisco M. Muñoz-Pérez,[1,2,*] Vicente Ferrando,[1] Walter D. Furlan,[3]
Juan C. Castro-Palacio,[1] J. Ricardo Arias-Gonzalez,[1] and Juan A. Monsoriu[1]

[1]Centro de Tecnologías Físicas, Universitat Politècnica de València, E- 46022, València, Spain.
[2]Laboratorio de Fibra Óptica, Universidad Politécnica de Tulancingo, División de Posgrado, Hidalgo, Mexico
[3]Departamento de Óptica, Universitat de València, Burjassot, E- 46100, Spain.
*Correspondence: fmmuope1@upvnet.upv.es

SUMMARY
The design and implementation of a multiplexed spiral phase mask in an experimental optical tweezer setup are presented. This diffractive optical element allows the generation of multiple concentric vortex beams with independent topological charges. The generalization of the phase mask for multiple concentric vortices is also shown. The design for a phase mask of two multiplexed vortices with different topological charges is developed. We experimentally show the transfer of angular momentum to the optically trapped microparticles by enabling orbiting dynamics around the optical axis independently within each vortex. The angular velocity of the confined particles versus the optical power in the focal region is also discussed for different combinations of topological charges.

Vortex; multiplexed vortices; phase mask; optical trapping; optical tweezers.

INTRODUCTION

Vortex dynamics is present in different scientific areas, such as gravitational vortices [1] or fluid vortices [2]. In optics, since the first proposal of an optical vortex in the late 1980s [3], multiple applications have been developed in optical trapping and manipulation, optical communication, encryption systems, and biosciences, among others [4-10]. In 1996, Gargaran et al showed the stable trapping of low refractive index microparticles through optical vortices [11]. With the development of new photonic technologies and components, the implementation of new techniques in vortex beams generation has grown, most notably with spiral phase plates [12]. Previous works have shown that spiral phase zone plates, also known as vortex lenses, allow the generation of a series of optical vortices distributed along the optical axis [9,13]. These vortex lenses are diffractive optical elements (DOEs) that can be easily generated with spatial light modulators (SMLs) [14,15].

The design of DOEs for vortex beam generation allows us to create multiple configurations, such as the multiplexed vortex phase mask proposed in this work. A vortex beam can be formed from a phase singularity characterized by its topological charge. These vortex beams present an angular momentum composed of an orbital component derived from the phase and intensity profile and a helical phase caused by their azimuthal phase dependence [16, 17]. The aforementioned characteristics make vortex beams potential tools in optical trapping systems. The incorporation of vortex DOEs into an optical tweezers system increases the flexibility and capacity for trapping and manipulating particles [18]. Optical vortices transfer angular momentum to trapped particles forcing them to move around the optical axis [19, 20], which is a valuable feature in an optical tweezers system. Furthermore, a spatially multiplexed vortex phase mask allows the generation of multiple simultaneous concentric optical vortices, each constituting a trapping and manipulation system. Previous works have demonstrated the utility of joint optical vortices as actuators in microfluidic and micromechanical systems [21, 22]. In short, the generation of multiplexed vortices through DOEs offers a viable option in the development of new optical tweezers systems and, thus, the generation of applications based on the maneuverability at the micro and nanoscales [23,24,]

In this work, we design and implement a new multiplexed vortex DOE in an experimental optical tweezers system. We present the phase mask profile design and numerical summations of the irradiance distribution. This DOE generates spatially multiplexed vortices. The intrinsic characteristics of the phase mask allow the multiple trapping of particles and the transfer of angular momentum to these particles by displacing them around the optical axis independently on each vortex. Experimental results show that multiplexed vortex beams generate stable dynamics in the trapping and displacement of the particles within the concentric rings.

EXPERIMENTAL PROCEDURES

*Material*
Multiplexed vortex lenses (MVL) for confined, rotational dynamics were set up in an optical tweezers design [14], as represented in Fig. 3. In short, a continuous wave laser beam ($\lambda$ = 1064 nm, Laser Quantum, Mod. Opus 1064) is incident on a half-wave plate ($\lambda/2$) followed by a linear polarizer ($P$), which lets the direction of the linear polarization of the beam to be set. The laser beam is guided by mirrors ($M1$ and $M2$) onto a x3 beam expander, as formed by lenses $L1$ and $L2$ (focal length $f_1$ = 50 mm and $f_2$ = 150 mm). The MVL is projected onto a spatial light modulator (SLM) display (Holoeye PLUTO-2.1-NIR-149, phase type, pixel size 8 $\mu m$ and resolution 1920 x 1080 pixels). A 1D blazed grating is added to each MVL, which purpose is to act as a linear phase carrier. It allows the diffracted light to be conducted towards the first order of diffraction, thus preventing noise caused by specular reflection from higher diffraction orders. The SLM is configured for a phase of $2.1\pi$ at a wavelength $\lambda$ = 1064 nm. The modulated MVL beam in the SLM is reduced by a $4f$ system consisting of $L_3$ ($f_3$ = 150 mm) and $L_4$ (focal length $f_4$ = 150 mm).

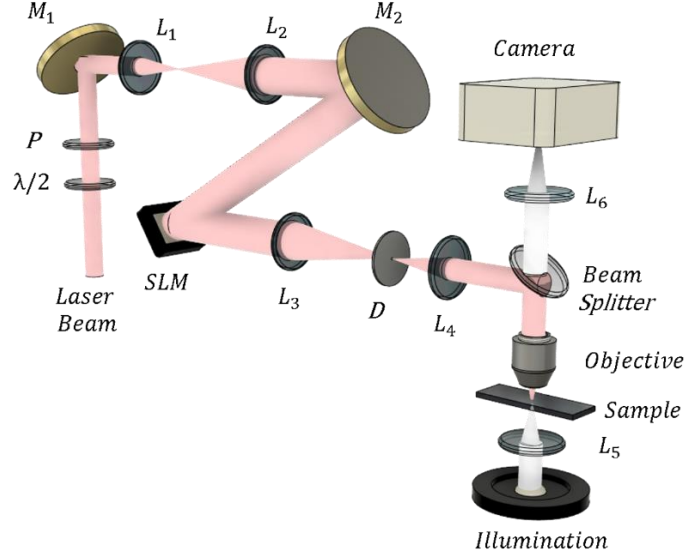

Figure 6. Experimental optical tweezers setup to confine particles by using MVLs.

The beam is spatially filtered by a diaphragm (**D**) placed at the focus of $L_3$, which lets only the first diffraction order pass. The SLM is tilted slightly to compensate for the added linear phase carrier, thus allowing the first-order diffraction to align with the optical axis of the diaphragm. The MVL image is then passed through either a 40X objective or a high-numerical aperture oil-immersion objective (Olympus UPLFLN 100X, NA= 1.3) used for rotational dynamics analysis., positioned at the focal plane $L_4$. The sample is illuminated with an LED light source (Thorlabs, Mounted High-Power, 1300 mA, Mod. MCWHL7) and the light collimated and focused on the sample through lens $L_5$ ($f_5$ = 30 mm). A beam splitter (BS) allows the transmission of the visible light from the sample through the rear focal plane of the objective. The BS prevents reflections of infrared light from being transmitted to the imaging system. The resulting image is focused with lens $L_6$ ($f_6$ = 50 mm). A CMOS camera (Edmund Optics, Mod. EO-10012C) is used for imaging purposes of the particles at the confining plane.

RESULTS AND DISCUSSION

Multiplexed spiral phase mask design
A spiral phase mask (SPM) is a DOE with a linear phase dependence only on the azimuthal angle. This phase distribution can be achieved by $\Phi(\theta_0) = mod_{2\pi}[m\,\theta_0]$, where $m$ is the so-called topological charge (an integer number different from zero), and $\theta_0$ is the azimuthal angle taken with respect to the optical axis of the pupil plane. Our strategy for generating multiplexed vortex beams, consists on integrating concentric SPMs with independent topological charges. In this regard, a multiplexed SPM (MSPM) is an arrangement of SPMs in concentric annular zones in a single DOE, which phase distribution $\tau(r,\theta_0)$ can be defined by:

$$\tau(r,\theta_0) = \begin{cases} mod_{2\pi}[m_1\theta_0], & 0 \leq r < r_1 \\ \ldots & \\ mod_{2\pi}[m_j\theta_0], & r_{j-1} \leq r < r_j \\ \ldots & \\ mod_{2\pi}[m_N\theta_0], & r_{N-1} \leq r < a \end{cases} \quad (1)$$

where $r$ is the radial coordinate, $a$ is the radius of the resulting DOE, $N$ is the total number multiplexed SPMs, and $m_j$ is the topological charge of the $j$-th SPM limited between radial distances $r_{j-1}$ and $r_j$, being $r_0 = 0$ and $r_N = a$.

Figure 1 shows the phase distribution of an MSPM ($N = 2$) with topological charges $m_1 = -7$ and $m_2 = 28$ used in this research work. The gray levels represent the phase modulation from 0 to $2\pi$. It is possible to observe the change of sign between the considered topological charges and the number of azimuthal periods corresponding to the value of each topological charge.

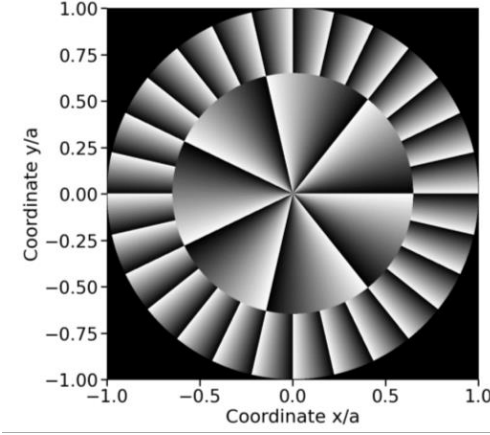

Figure 1. Phase distributions of a MSPM ($N=2$, $m_1 = -7$ and $m_2 = 28$).

Focusing properties with MSPM
The irradiance of the DOE is provided by the transmittance, $t(\zeta, \theta_0) = q(\zeta)\exp[im\theta_0]$, of the diffractive element when it is illuminated by a plane wave of wavelength $\lambda$, where $\zeta = (r/a)^2$. Let us now consider an MSPM with a phase distribution given by Eq. 1 placed at the exit pupil of a microscope objective. The resulting irradiance within the Fresnel approximation as a function of the axial distance from the pupil plane $z$ is [27]:

$$I(u,v,\theta) = u^2 \left| \sum_{j=1}^{N} \int_{\zeta_{j-1}}^{\zeta_j} \int_0^{2\pi} q(\zeta) \exp(im_j\theta_0) \exp(-i2\pi u\zeta) \exp[i4\pi uv\zeta^{1/2} \cos(\theta - \theta_0)] \, d\zeta d\theta_0 \right|^2, \quad (2)$$

where $u = a^2/2\lambda z$ is the reduced axial coordinate, $v = r/a$ is the normalized transverse coordinate, and $\theta$ is the azimuthal coordinate. By solving the integral for the angular dependence, we find:

$$\int_0^{2\pi} \exp[im_j\theta_0] \exp[i4\pi uv\zeta^{1/2} \cos(\theta - \theta_0)] d\theta = 2\pi \exp[im_j(\theta_0 + \pi/2)] J_{m_j}(4\pi uv\zeta^{1/2}), \quad (3)$$

being $J_{m_j}$ the Bessel function of the first kind of order $m_j$. As a consequence, Eq. 2 reduces to

$$I(u,v) = 4\pi^2 u^2 \left| \sum_{j=1}^{N} \int_{\zeta_{j-1}}^{\zeta_j} q(\zeta) \exp(-i2\pi u\zeta) J_{m_j}(i4\pi uv\zeta^{1/2}) d\zeta \right|^2, \quad (4)$$

We have computed the irradiances provided with the MSPM shown in Fig. 1 by using Eq. (4). The result can be seen in Fig. 2, where it is possible to discern two main vortices formed at the focal plane, both exhibiting a cosine azimuthal dependence. The interference intensity profile stems from the superposition of both vortices as a consequence of the difference in topological charges. The intensity map of the inner vortex is homogeneous at the transverse, focal plane (see Fig. 2C), but the outer vortex is constituted of angularly-distributed lobes associated to alternating constructive and destructive wave superpositions. The total number of lobes in the outer vortex is 35, which is congruent with the relationship $(m_2 - m_1) = (28 + 7) = 35$ derived elsewhere [19].

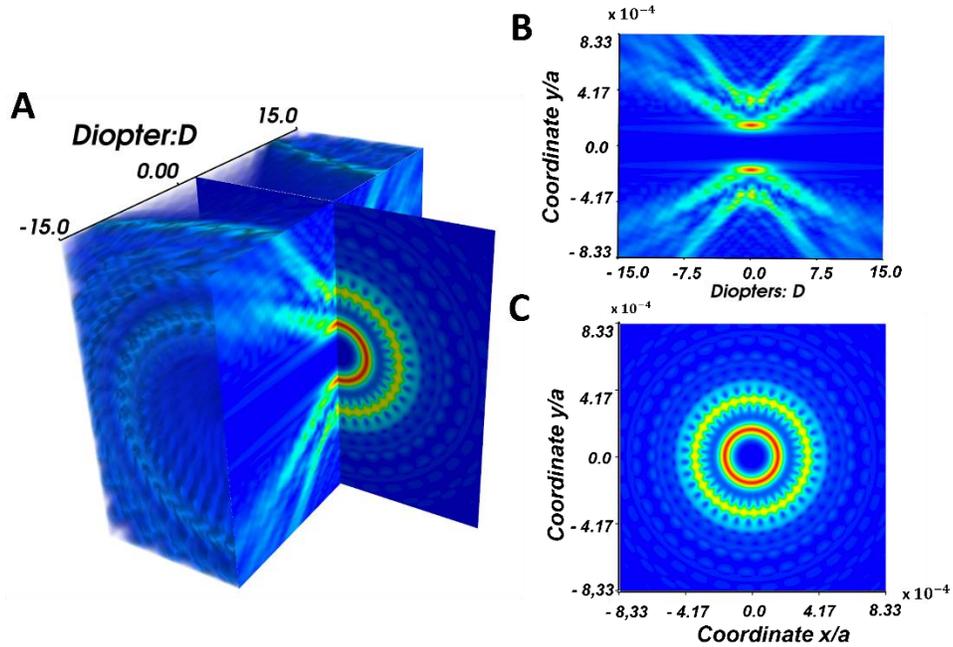

Figure 2. (A) Computed irradiance with the MSPM shown in Fig. 1. (B) and (C) Longitudinal and transversal irradiances, respectively, at the indicated planes.

Trapping and manipulation with a MSPM

We present the experimental manipulation of microparticles through the multiplexed vortex beams shown in Fig. 2. Our optical setup allows the trapping of several microparticles arranged in two circular rings with nearly independent dynamics. In this regard, Fig. 3 shows the steady rotational motion of serial polystyrene beads (diameter ~2 µm) in each vortex. The arrows indicate the direction of particle rotation, consistent with the sign of the topological charges herein used. In particular, since $m_1$ is negative, the motion of the particles is levorotatory (anti-clockwise, as shown in the depicted planes of Fig. 3). The opposite takes place for the outer vortex, where $m_2$ is positive, making the dynamics dextrorotatory (clockwise in Fig. 3).

An individual particle was selected in each concentric vortex for analyzing further the rotational motion in the fluid: the one in the inner vortex is marked with a blue dot, whereas that on the outer vortex with a red dot.

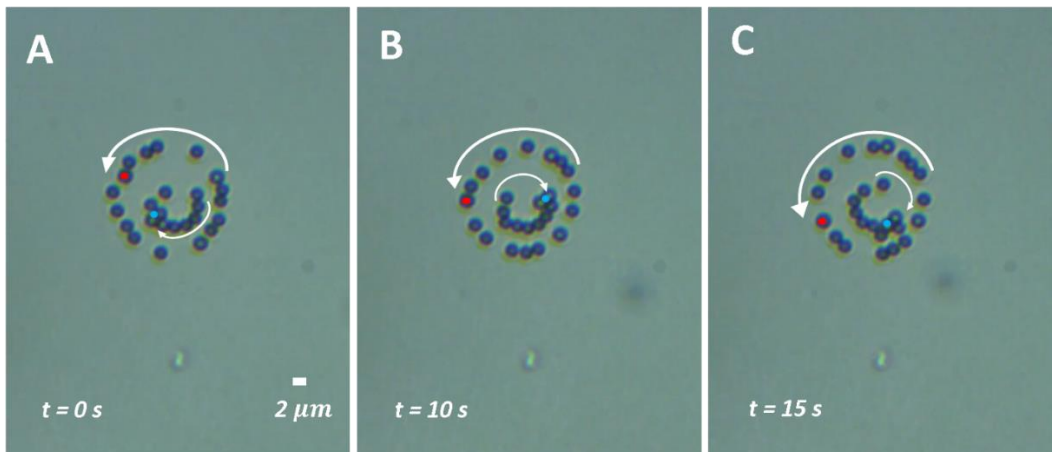

Figure 3. Sequence of images (A-C) for the dynamic trapping of microparticles around MSPM-based optical tweezers (see Visualization 1). The inner vortex has a topological charge $m_1 = -7$ and the outer one $m_2 = 28$.

An essential feature of our MSPM-based optical tweezers is the opposite spin directions between vortices, which are controlled by the topological charges of the phase mask. In this regard, by flipping signs in $m_1$ and $m_2$, the two rotational motions of the particles reverse, as observed in Fig. 4(A-C). As expected, If the signs are the same for both topological charges, the direction of rotation direction becomes the same for both vortices.

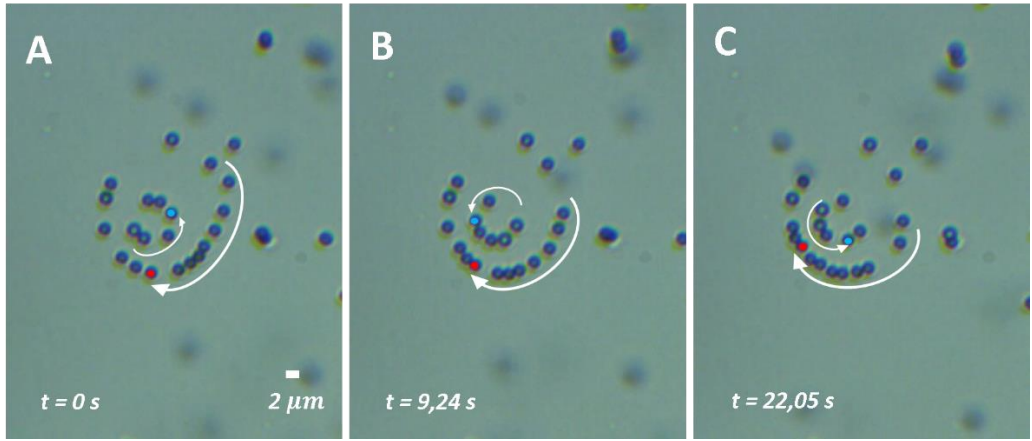

Figure 4. Sequence of images (A-C) for the dynamic trapping of microparticles around MSPM-based optical tweezers (see Visualization 2). The inner vortex has a topological charge $m_1 = 7$ and the outer one $m_2 = -28$.

The steady motion of the microparticles is a consequence of the torque exerted by light, which transfers angular momentum to the trapped particles, counteracted by the friction in the fluid environment. The angular velocity, $\omega$, of the orbiting dynamics depends on the confining vortex. To demonstrate this, we performed a more detailed analysis with a higher magnification objective (100x, NA = 1.3). The results are displayed in Fig. 5, where it is observed generally that the angular velocity of the external vortex is lower than that of the internal vortex.

The behavior of the angular velocity as a function of the optical power measured at the focal plane is shown in panels (A) and (B) for different $m_1$ and constant topological load $m_2$. The angular velocity for the internal vortex increases with the optical power, unlike for the external vortex, which remains constant within experimental error.

With regards to variations in the topological charge, Fig. 5 (C), it is observed that the rotational motion of the particles in the internal vortex starts at a threshold topological charge right before $m_1 = 4$ and that their angular velocity exhibit a maximum for $m_1 = 7$. The diameter of the internal vortex is smaller than that of the particles for $m_1 \leq 4$, making the vortex approach a point trap, hence only capable of confining a single particle in the vortex center. The transferred angular momentum may spin the particle around its center but it is insufficient to make it observable. When $m_1 > 7$, the diameter of the internal vortex increases, causing a redistribution of the light power on a larger area at the focal region, which in turn reduces the irradiance and the angular momentum imparted by the photon flux onto the particles. These results are consistent with those of Liang et al. [28], who studied the behavior of particles within a single vortex. The particles in the outer vortex, in contrast, although they exhibit motion at low values of $m_1$ (see Fig. 5 (C)), their angular velocity increases only slightly with $m_1$, which is also due to the power distribution on a larger area for this external vortex.

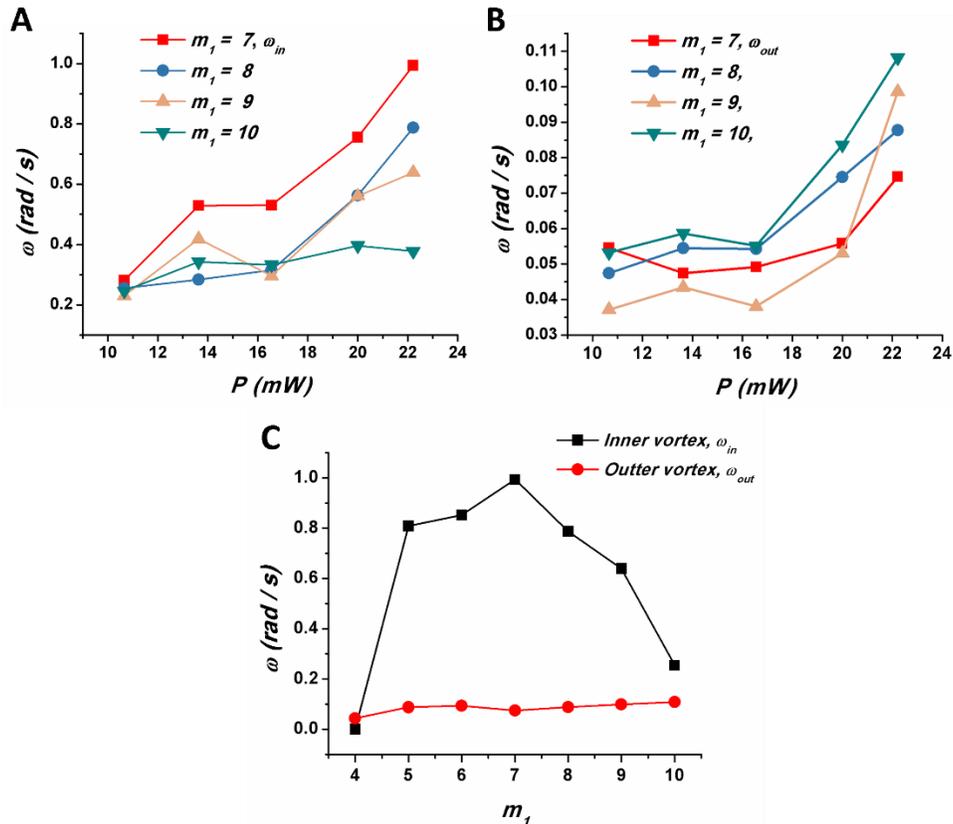

Figure 5. Orbiting dynamics of microparticles in an MSPM. (A-B) Modulus of the angular velocity for particles confined in the inner ($\omega_{in}$) and outer ($\omega_{out}$) vortices vs. optical power at the focal region for different topological charges of the internal vortex and $m_2 = -38$. (B) Angular velocity for particles confined in the inner and outer vortices vs. topological charge of the internal vortex at a focal power of 22.21 mW.

Conclusions

An MSPM has been designed and implemented on an experimental arrangement of optical tweezers, which allows the trapping and manipulation of particles on independent vortices. A diffractive structure enables the formation of an internal vortex with a topological charge $m_1$ and an external vortex with a topological charge $m_2$, both at the focal plane. The trapped particles orbit around the vortex centers due to the phase distribution design, which enables angular momentum transfer. The orbiting direction is dependent on the topological charge sign, with which it is possible to model each individual vortex. The MSPM exhibits stable trapping capacity for multiple particles in each vortex, regardless of the spin direction. It also allows an independent rotation control between vortices, opening a away to trapping microstructures with dynamic manipulation. We believe that our MSPM can be implemented to control microrobots, given its capacity to transduce light power into mechanical work.

SUPPLEMENTAL INFORMATION
Video S1. MVL_1: Dynamic of trapped microparticles with MSPM-based optical tweezers. The inner vortex has a topological charge $m_1 = -7$ and the outer one $m_2 = 28$.
Video S2. MVL_2: Dynamic of trapped microparticles with MSPM-based optical tweezers. The inner vortex has a topological charge $m_1 = 7$ and the outer one $m_2 = -28$.

ACKNOWLEDGMENTS
This work was supported by the Spanish Ministerio de Ciencia e Innovación (grant PID2019-107391RB-I00) and by Generalitat Valenciana (grant PROMETEO/2019/048), Spain. F.M.M.P also acknowledgment the financial support from the Universitat Politècnica de València (PAID-01-20-25), Spain.

AUTHOR CONTRIBUTIONS
F.M.M.P., J.R.A.G., and J.A.M. conceived and designed the experiments. F.M.M.P. and V.F. performed the experiments. W.D.F. and J.C.C.P analyzed the data. F.M.M.P. wrote the manuscript.

DECLARATION OF INTERESTS
The authors declare no competing interests.